\documentclass[prd,10pt,tightenlines,aps,letterpaper,amsmath,amssymb,preprintnumbers,
               showpacs,floatfix,longbibliography,nofootinbib,superscriptaddress,twocolumn]{revtex4-2}
\usepackage{subfiles}
\usepackage{siunitx}
\usepackage{graphicx}
\graphicspath{ {./figures/} }
\usepackage[export]{adjustbox}
\usepackage[caption=false]{subfig}
\usepackage{bm}  
\usepackage{makecell} 
\allowdisplaybreaks 
\usepackage[hypertexnames=true]{hyperref}

\usepackage[capitalize]{cleveref}
\crefname{equation}{Eq.}{Eqs.}
\crefname{figure}{Fig.}{Figs.}
\crefname{table}{Tab.}{Tabs.}
\crefname{section}{Sec.}{Secs.}
\crefname{appendix}{App.}{Apps.}
\Crefname{table}{Table}{Tables}
\Crefname{figure}{Figure}{Figures}

\hypersetup{
    colorlinks=true,       
    linkcolor=blue,          
    citecolor=blue,        
    filecolor=blue,      
    urlcolor=blue           
}

\usepackage{slashed}
\usepackage{mathtools}
\usepackage{listings}
\usepackage{pgf}
\usepackage{fancyvrb}
\usepackage{longtable}
\usepackage{siunitx}
\usepackage{braket}
\usepackage{duckuments}
\usepackage{soul}
\usepackage{comment}

\usepackage{enumitem} 

\usepackage[normalem]{ulem} 
\usepackage{dsfont} 

\usepackage{tikz}
\usetikzlibrary{shapes,arrows,chains}
\usetikzlibrary{calc}
\usetikzlibrary{patterns}
\usetikzlibrary{patterns}
\usetikzlibrary{positioning,decorations.pathmorphing,decorations.markings}

\tikzset{>=latex}
\tikzset{every picture/.style={line width=0.75pt}}

\usepackage{array}
\newcolumntype{P}[1]{>{\centering\arraybackslash}p{#1}}

\usepackage{soul}

\DeclareMathAlphabet{\mathbbb}{U}{bbold}{m}{n}

\definecolor{listinggreen}{rgb}{0,0.6,0}
\definecolor{listinggray}{rgb}{0.5,0.5,0.5}
\definecolor{listingmauve}{rgb}{0.58,0,0.82}
\definecolor{listingkeywordcolor}{rgb}{1.0,0.4,0.0}
\definecolor{listinglightgray}{rgb}{0.8863,0.8863,0.8863}

\usepackage{colortbl}

\newcommand{\Egap}{\hat{E}_{\text{miss}}}

\newcommand{\nn}{I=1}
\newcommand{\deut}{I=0}
\newcommand{\nprop}{$2.12\times 10^7$}
\newcommand{\nconf}{$16,368$}

\begin{document}

\begin{abstract}
Lattice QCD has historically produced energy results interpretable as either estimates relying on implicit assumptions about asymptotic behavior or one-sided upper bounds. New Lanczos methods providing two-sided bounds with less-restrictive assumptions are introduced and quantified in a high-statistics calculation with unphysical quark masses. 
Two-sided bounds without spectral assumptions provide sub-percent constraints on the nucleon mass. 
Other bounds, which assume all states in a given energy window are resolved, provide meaningful two-sided constraints on nucleon-nucleon scattering phase shifts.
\end{abstract}

\preprint{FERMILAB-PUB-26-0040-T}
\preprint{MIT-CTP/5997}

\title{Excited-state uncertainties in lattice-QCD calculations of hadron masses and scattering phase shifts}
\author{William Detmold}
\affiliation{Center for Theoretical Physics - A Leinweber Institute, Massachusetts Institute of Technology, Cambridge, MA 02139, USA}
\author{Anthony V. Grebe}
\affiliation{Department of Physics and Maryland Center for Fundamental Physics, University of Maryland, College Park, MD 20742, USA}
\affiliation{Fermi National Accelerator Laboratory, Batavia, IL 60510, USA}
 \author{Daniel C. Hackett}
\affiliation{Fermi National Accelerator Laboratory, Batavia, IL 60510, USA}
\author{Marc~Illa}
\affiliation{InQubator for Quantum Simulation (IQuS), Department of Physics, University of Washington, Seattle, WA 98195, USA}
\affiliation{Physical Sciences Division, Pacific Northwest National Laboratory, Richland, WA 99354, USA}
\author{Robert J. Perry}
\affiliation{Center for Theoretical Physics - A Leinweber Institute, Massachusetts Institute of Technology, Cambridge, MA 02139, USA}
\author{Phiala E. Shanahan}
\affiliation{Center for Theoretical Physics - A Leinweber Institute, Massachusetts Institute of Technology, Cambridge, MA 02139, USA}
\author{Michael L. Wagman}
\affiliation{Fermi National Accelerator Laboratory, Batavia, IL 60510, USA}
\collaboration{NPLQCD Collaboration}
\begin{figure}
  \vskip -1.0cm
  \leftline{\includegraphics[width=0.15\textwidth]{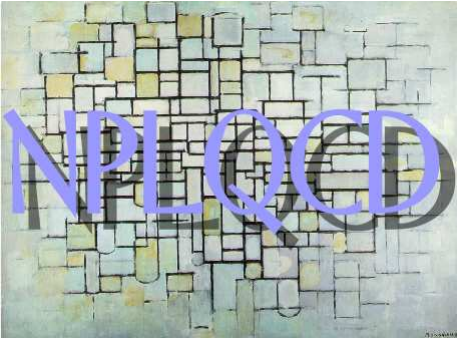}}
\end{figure}

\maketitle

{\it Introduction:} Lattice regularization of quantum field theory (LQFT) provides a well-defined computational pathway for the evaluation of physical quantities, with theoretical control over all aspects of the calculation. 
In lattice quantum chromodynamics (LQCD) in particular, this has led to calculations in many phenomenologically important areas including the structure and interactions of hadrons, and the nature of strongly interacting matter in extreme conditions. LQCD calculations are also instrumental in determining the parameters of the Standard Model and in searching for physics beyond it.
Nevertheless, in practical calculations, the  limits needed to guarantee correctness of the field theory predictions are not realized numerically, leading to systematic uncertainties that must be quantified. Prominent examples of such uncertainties are those associated with the extrapolation of LQCD calculations to the continuum and infinite volume limits. When numerical LQCD calculations are performed in the regime where the lattice spacing is sufficiently small and the lattice volume is sufficiently large, theoretical understanding of discretization and finite-volume effects allows for well-motivated extrapolation to the necessary limits. 

This work focuses on an equally important source of uncertainty\footnote{While different sources of uncertainties do not completely decouple, the discussion here ignores such interactions for simplicity.}  that arises in  extracting physical quantities such as hadron masses from $n$-point Euclidean correlation functions (correlators) built from field operators.
In specific examples below, two-point correlators 
\begin{equation}
  C_{ab}(t)={\rm Tr}[T^{L_t}{\cal O}_a(t){\cal O}_b^\dagger(0)] \, / \, {\rm Tr}[T^{L_t}]  
\end{equation}
are considered, where $T$ is the transfer matrix,  ${\cal O}_{a,b}$ are linear combinations of products of local field operators, and $L_t$ is the temporal extent of the lattice geometry under thermal field theory boundary conditions.
These correlators, and their generalizations to higher-point functions, are simply related to the energy eigenvalues and matrix elements in the  $t\rightarrow \infty$ limit of infinite temporal separation of the field operators, but at any finite separation are contaminated by a formally infinite set of excited states.
While this limit provides an exactness guarantee,  for practical calculations it is important to understand what level of control exists for the corresponding systematic uncertainties on quantities extracted at finite time, generically referred to as ``excited-state effects.''
More details can be found in the companion work of Ref.~\cite{long}.

For stable hadrons, correlators can usually be computed with statistical control at imaginary times $t$ that are much larger than the inverse of the smallest excitation energy $\delta$. In this case, excited-state contributions to correlators are suppressed by $e^{-\delta t}\ll1$ relative to ground-state contributions, and it is therefore possible to model the correlators accurately as ground-state contributions plus small corrections. 
Nevertheless, even in the single-hadron sector, there are situations in which even for $e^{-\delta t}\ll1$, prefactors on these terms are large and excited states are hard to control in practical calculations. One example of this situation can be found in precision determinations of nucleon axial form factors~\cite{Bar:2018xyi,Jang:2019vkm,RQCD:2019jai,Park:2021ypf,Jang:2023zts,Alexandrou:2024tin,Hall:2025ytt,Barca:2025det}.

For multi-hadron systems, spectroscopy
is fundamentally more challenging for two reasons. Firstly, it is generically the case that the signal-to-noise ratios for Monte-Carlo estimates of multi-hadron correlators  decay rapidly in extensive parameters such as the temporal separation of field operators, making the $t\to\infty$ limit inaccessible in practical calculations. Secondly, 
multi-hadron systems necessarily present energy gaps $\delta$ that approach zero as the lattice volume increases. Consequently, statistical control of correlators at $t \gg 1/\delta$ is not achievable with available computational resources and $e^{-\delta t}$ is not a small parameter. In contrast to lattice spacing and volume effects, systematic uncertainties from excited states cannot be treated as small corrections in state-of-the-art calculations for these systems. Given these difficulties, spectroscopic information must either be obtained from methods that are robust to these effects or be based on assumptions about them.

{\it Assumptions in LQCD spectroscopy:}
To date, the extraction of hadronic energies from (sets of) correlators has relied on an often-implicit assumption, namely:
\begin{itemize}[leftmargin=*]
\item[] \textit{$N$-state-saturation assumption}: There exists a region of Euclidean time in which the correlator (matrix) is accurately described by a truncated spectral decomposition with $M \geq N$ states, where $N$ states approximate genuine energy eigenstates up to statistical noise (e.g., the $N$ lowest-energy states).
\end{itemize}
The validity of such an $N$-state-saturation assumption relies on the identification of interpolating operators that have sufficiently large overlaps with all the state(s) of interest, and sufficiently small overlaps with all other states. The $N=1$ case is often an assumption of ground-state saturation. 

As discussed in Ref.~\cite{long}, there are multiple examples of  calculations in LQCD, and in solvable field theories, in which this type of assumption is clearly not valid. Symmetric correlators (in which $\mathcal{O}_a=\mathcal{O}_b$) do not avoid this issue, and although asymmetric correlators have additional failure modes~\cite{long}, analyses of both can, and do, fail in real examples. 
For example, 1) in the $I=1$ vector channel, Ref.~\cite{Dudek:2012xn} showed that the presence or absence of ``$\pi\pi$'' or ``$\rho$'' interpolating operators dramatically alters the spectrum; 2) in the two-nucleon channels, Ref.~\cite{Amarasinghe:2021lqa} showed extreme operator dependence of the extracted spectrum from consideration of different combinations of dibaryon operators; and 3) in a solvable field theory, the companion paper \cite{long} shows discrepancies between energy estimators and the exact spectrum at very high significance. 
These failures highlight the importance of tools for correlator analysis based on weaker assumptions. 

In this work, three such tools are discussed and their effectiveness is quantified in the context of a high-statistics LQCD calculations of one- and two-nucleon correlators. Based on these tools, this work outlines a more robust approach to LQCD correlator analysis in regimes where excited states are not parametrically suppressed by sufficiently large temporal separations.

As is well-known in LQCD \cite{Wilson:1981,Berg:1981zb,Falcioni:1981mm,Michael:1982gb,Fernandez:1987ph,Luscher:1990ck}, symmetric (matrices of) correlators, defined by the use of identical (sets of) interpolating operators at the source and sink, provide one-sided \emph{variational bounds} on energy eigenvalues that only assume that energy eigenvalues are real and thermal effects are negligible. These bounds stem from the convex nature of the correlator (matrix). While variational bounds provide meaningful constraints on the locations of LQCD energy eigenvalues, they formally do not provide two-sided bounds required to constrain multi-hadron scattering amplitudes in phenomenologically useful ways without additional $N$-state-saturation assumptions.

In recent work that applies Lanczos methods to correlator analysis, \emph{residual bounds} that provide two-sided bounds on energy eigenvalues have been introduced in the LQFT context~\cite{Wagman:2024rid}. These bounds do not require any assumptions beyond Hermiticity (and, in some uses, state labeling~\cite{long}). Their size depends on 
how well the infinite-dimensional transfer matrix is approximated in the relevant energy regime by the finite-dimensional projections of that operator that are accessible from a set of computed correlators.
Crucially, correlator data provides a way to assess this quality of approximation 
through the calculation of the residual-norm-square $B$ discussed further below.
Here, it is seen that two-sided residual bounds constructed from $B$ provide strong constraints on energies in one- and two-nucleon systems.

Two-sided \emph{gap bounds} are further introduced in this work. 
Their utility relies on the smallness of the same parameter $B$, but gap bounds
 are parametrically tighter than residual bounds. Gap bounds rely on a stronger assumption than residual bounds:
\begin{itemize}[leftmargin=*]
\item[] \textit{No-missing-states assumption}: There are exactly $N$ states with energy less
than a specified threshold $\Egap$ that make non-zero contributions to a correlator (matrix), neglecting thermal effects. 
\end{itemize}
As with $N$-state saturation, the validity of gap bounds requires an explicit assumption about the energy spectrum. However, no assumptions are made about the sizes of interpolating operator overlaps with the $N$ states below $\Egap$, making gap bounds more rigorous than energy estimators based on $N$-state saturation. For calculations very close to the physical values of the quark masses and near to the continuum, infinite-volume limit, this is particularly pertinent as experimental information can be used to constrain the energy spectrum but not to constrain overlaps.
When no-missing-states assumptions are valid, gap bounds may provide useful systematic uncertainty estimates even when $e^{-\delta t}$ is not small. 
Nevertheless, gap bounds will be violated if $\Egap$ is poorly estimated, so care should be taken in their application. Below, the utility of gap bounds  in extracting nucleon-nucleon interactions from two-nucleon correlators is investigated and found to provide phenomenologically useful constraints.

Spectroscopy under two-sided constraints is a paradigmatic improvement over previous approaches and is \textit{required} for quantifying excited-state systematic uncertainties when $e^{-\delta t}$ is not small. Much focus in spectroscopy has been on careful construction of physically motivated operators with overlap onto given states of interest, as well as techniques such as generalized eigenvalue problem (GEVP)~\cite{Fernandez:1987ph,Luscher:1990ck,Blossier:2009kd} and Prony-Ritz~\cite{Fleming:2004hs,Lin:2007iq,Beane:2009kya,Beane:2009gs,Fleming:2009wb,Beane:2009kya,Beane:2009gs,Aubin:2010jc,Aubin:2011zz,Fleming:2023zml,Wagman:2024rid,Hackett:2024xnx,Ostmeyer:2024qgu,Chakraborty:2024scw,Hackett:2024nbe,Abbott:2025yhm,Ostmeyer:2025igc,Tsuji:2025zdn} energy estimators to maximize the information that can be extracted from sets of correlators. However, improved interpolating-operator and energy-estimator construction does not address the need to quantify systematic uncertainties from excited-state effects. In contrast, the two-sided bounds available in the Lanczos approach quantify these effects and lead to improved rigor in practical LQCD calculations.

{\it Lanczos methodology and two-sided bounds:}
The Lanczos method as applied in LQCD \cite{Wagman:2024rid,Hackett:2024xnx,Ostmeyer:2024qgu,Chakraborty:2024scw,Hackett:2024nbe,Abbott:2025yhm,Ostmeyer:2025igc,Tsuji:2025zdn} seeks to constrain the spectrum of the Euclidean-space transfer matrix from correlators (scalar Lanczos) or from matrices of correlators (block Lanczos).  
Defining the true eigenvalues and eigenvectors of the transfer matrix, and the associated energy eigenvalues, as 
\begin{equation}
T\ket{n} = \lambda_n \ket{n},\qquad {\rm with} \quad
E_n = -\ln\lambda_n \ ,
\end{equation}
the Lanczos algorithm iteratively constructs left and right Ritz vectors, $\ket{y^{L/R(m)}_k}$, and Ritz values, $\lambda^{(m)}_k$, from correlator matrices. From them, the $m^{\rm th}$-iteration approximation to the transfer matrix can be built as 
\begin{equation}
    T^{(m)} = \sum_{k=0}^{rm-1} \ket{y^{R(m)}_k} \lambda^{(m)}_k \bra{y^{L(m)}_k} ~ 
\end{equation}
for a rank-$r$ correlator matrix.

Residual norm-squares that quantify the differences between accessible projections of the true and Lanczos-approximated transfer matrix can formally be defined as 
\begin{equation}
    B^{R/L(m)}_k \equiv  
     \frac{ \braket{y^{R/L(m)}_k | [T - T^{(m)}]^\dagger [T-T^{(m)}] | y^{R/L(m)}_k} }{ \braket{y^{R/L(m)}_k | y^{R/L(m)}_k} } ~ ;
     \label{eq:Bdef}
\end{equation}
see Ref.~\cite{long} for details on practical implementation of \cref{eq:Bdef} using the filtered Rayleigh-Ritz perspective \cite{Abbott:2025yhm} and outlier-robust bootstrap-median estimators.

In terms of these quantities, the residual bounds \cite{Parlett} on energy eigenvalues are given by the range
\begin{equation}
\begin{split}
   E_n &\in \left[ -\ln\left(\lambda_k^{(m)} + \sqrt{B_k^{R/L(m)}}\right), \right. \\
   &\hspace{20pt} \left. \  -\ln\left(\lambda_k^{(m)} - \sqrt{B_k^{R/L(m)}}\right) \right]
   \label{eq:E_window}
   \end{split}
\end{equation}
in which one or more true energy eigenvalues is guaranteed to exist. It is important to note that residual bounds do not guarantee that an interval contains any \emph{particular} eigenvalue such as~the ground state $E_0$. 
They also do not imply statements about regions in which eigenvalues are guaranteed not to exist. 
Unlike variational bounds, residual bounds remain valid at non-zero temperature.

Gap bounds on energies are similarly given by
\begin{equation}
\begin{split}
   E_n &\in \left[ -\ln\left(\lambda_k^{(m)} + \frac{B_k^{R/L(m)}}{G_k^{(m)}} \right), \right. \\
   &\hspace{20pt} \left. \  -\ln\left(\lambda_k^{(m)} - \frac{B_k^{R/L(m)}}{G_k^{(m)}} \right) \right],
   \label{eq:E_window_Gap}
   \end{split}
\end{equation}
where $G_k^{(m)}$ is a gap parameter that will be discussed below. 
Since gap bounds depend on $B_k^{R/L(m)}$ rather than $\sqrt{B_k^{R/L(m)}}$, they are parametrically more constraining than residual bounds as the residual norm-square decreases. 
A range of gap bounds have been derived that differ in the definition of the gap parameter. For the original  Davis-Kahan bounds \cite{Davis:1970,Parlett}, it is defined as 
\begin{equation}\label{eq:gkdef}
   G_k^{(m)} \equiv \min_{\lambda_{n'} \in \{\lambda_n \neq \lambda\}} \left|  \lambda_{n'} - \lambda_k^{(m)} \right|,
\end{equation}
where  $\lambda$ is the closest eigenvalue to $\lambda_k^{(m)}$.
Here, $G_k^{(m)}$ is the gap between the Ritz value $\lambda^{(m)}_k$ and the second-closest eigenvalue of $T$ to $\lambda^{(m)}_k$.
For Haas-Nakatsukasa gap bounds \cite{Haas:2025}, which have been recently introduced for the case of block Lanczos and are more constraining, 
the gap is replaced by the distance between the Ritz value and the closest eigenvalue not in one-to-one correspondence with a Ritz vector in the block. Under a no-missing-states assumption, $G_k^{(m)}$ can be estimated as 
\begin{equation}\label{eq:GkEgap}
  \hat{G}_k^{(m)} \equiv \left| e^{-\Egap} - \lambda_k^{(m)} \right|,
\end{equation}
where $\Egap$ is an estimate of the energy of the lowest-energy state not adequately approximated by some interpolating operator.

\begin{table*}[!t]
    \centering
    \begin{ruledtabular}
    \begin{tabular}{clllll}
    State & $\overline{E}_0$  & $\delta \overline{E}_0$ & Gap & Residual & $\Egap$
    \\
    \hline
       $N$               & 1.20349 & 0.00018 & 0.00027 & 0.0078  & $\overline{E}_1 = 1.782$ \\
      $NN\ \nn, \ 1D$         & 2.40311 & 0.00043 & 0.0029 & 0.013  & $\hat{E}_1(M_N, M_N)=  2.463$\\
      $NN\ \deut, \ 1D$       & 2.40207 & 0.00041 & 0.0026 & 0.012 & $\hat{E}_1(M_N, M_N)=  2.463$\\\hline

     $NN\ \nn, \ 3D$   & 2.40252 & 0.00041 & 0.0021 & 0.017 & 
     $\hat{E}_3(M_N, M_N)=  2.572$\\
      $NN\ \deut,\ 10D$ & 2.40094 & 0.00041 & 0.0025 & 0.026 & 
      $\hat{E}_0(M_\Delta, M_\Delta)=2.664$ \\
     \end{tabular}
    \end{ruledtabular}
    \caption{Lanczos ground-state  results in units in which the lattice spacing $a=1$. $\overline{E}_0$ is the large-$m$ median Lanczos energy estimator and $\delta\overline{E}_0$ is its statistical uncertainty, ``Gap'' labels the  lower gap bound widths at $68\%$ confidence, and ``Residual'' labels the residual bound widths at 68\% confidence. 
    Scalar Lanczos results (top three rows) use Davis-Kahan gap bounds, while block Lanczos results (bottom two rows) use Haas-Nakatsukasa gap bounds and are from 3(10) dimensional operator sets for $I=1$(0). Also shown are the $\Egap$ values used in constructing gap bounds for each case. For the two-nucleon channels, $\
    \hat{E}_{n}(M_1, M_{2}) = \sqrt{ M_1^2 + n \left( \frac{2\pi }{L} \right)^2 } + \sqrt{ M_{2}^2 + n \left( \frac{2\pi }{L} \right)^2 }$ are the non-interacting energy levels of hadrons of mass $M_1$ and $M_2$. 
    \label{tab:results}
    }
\end{table*}

In Ref.~\cite{long}, further details of the Lanczos method and of both types of bounds are presented, along with an extensive exploration of the calibration of the resulting uncertainties in a solvable field theory. 

{\it Lattice QCD calculations:} To demonstrate the phenomenological value in these newly developed bounds, a high-statistics LQCD calculation using a single ensemble of gauge field configurations is performed at quark masses corresponding to a pseudoscalar meson mass of $m_\pi\sim 800$ MeV. The calculation uses \nprop\ sparsened quark propagators calculated on \nconf\  configurations and further details  are discussed in Ref.~\cite{long}.

Scalar Lanczos analysis, along with Bayesian model-averaged~\cite{Jay:2020jkz} (BMA) fits to GEVP principal correlators~\cite{Fernandez:1987ph,Luscher:1990ck,Blossier:2009kd}, is applied to a single symmetric nucleon correlator constructed from smeared quark fields. The resulting energy estimator, statistical uncertainty, and both gap and residual bounds are presented in \cref{tab:results}. The statistical uncertainty is 0.02\%, while the gap and residual bound uncertainties are 0.03\% and 0.6\%, respectively.
Even using residual bounds that only assume Hermiticity of $T$, these uncertainties are at the level of or smaller than other systematic effects such as the continuum extrapolation, isospin breaking corrections, and QED effects. Residual bounds therefore provide sufficiently precise control over the determination of energy eigenvalues  such that assumptions on the spectrum and interpolating operator overlaps are not needed in practice to determine the nucleon mass at these quark masses.\footnote{It is however an assumption that the lowest-energy state identified corresponds to the ground state of the system.} 

\begin{figure}[t]
    \centering
    \includegraphics[width=0.95\columnwidth]{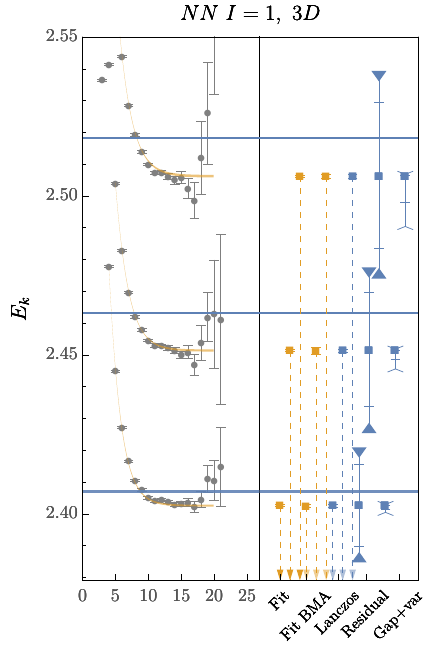}
    \caption{Effective masses of GEVP principal correlators compared with highest-weight fit (orange band, left, and first point, right) and Bayesian model averaged~\cite{Jay:2020jkz} (BMA) multi-state fit results (second orange point, right). Median averages of large-$m$ Lanczos energy estimators, residual bounds, and combined gap and variational bounds are shown as blue points, right. Arrows denote one-sided variational bounds. Horizontal lines denote bootstrap medians of two-sided bounds; triangles (chevrons) denote 68\% confidence intervals for residual (gap plus variational) bounds.}
    \label{fig:nn_summary}
\end{figure}

The $I=0$ and $I=1$ two-nucleon systems are studied with the same analysis techniques based on correlator matrices built from dibaryon and hexaquark operators (in the $I=1$ ($I=0$) channel, up to 8 (21) operators are used, see Ref.~\cite{long} for details). Results of the analysis for both the Lanczos method and BMA fits to GEVP principal correlators are shown for the lowest-energy states in the $I=1$ channel in Fig.~\ref{fig:nn_summary}. Results for both scalar and block Lanczos are given in \cref{tab:results}. The relative accuracy of these results for the $I=0$ channel is 0.02\% for statistical uncertainties, 0.1\% for gap bounds, and 0.5\% for the most constraining (scalar) residual bounds. 

Although the residual bounds are impressively tight, the gaps between near-threshold scattering states in this finite volume system are at the level of 0.3\% of the threshold energy, and potential bound states may have energy gaps that are considerably smaller than this.  Determination of scattering phase shifts from FV energies requires resolving such gaps, so at the currently accessible level of statistical precision, residual bounds are insufficient to constrain scattering phase shifts.
With additional sampling and improved interpolating operator constructions, it is realistic to think that residual bounds can be reduced by a factor of a few and a spectral-assumption-free path to information about hadronic interactions exists.
Nevertheless, at present, determination of hadronic interactions must be based on assumptions even at these unphysically heavy quark masses. 

\begin{figure}
    \centering
    \includegraphics[width=0.95\linewidth]{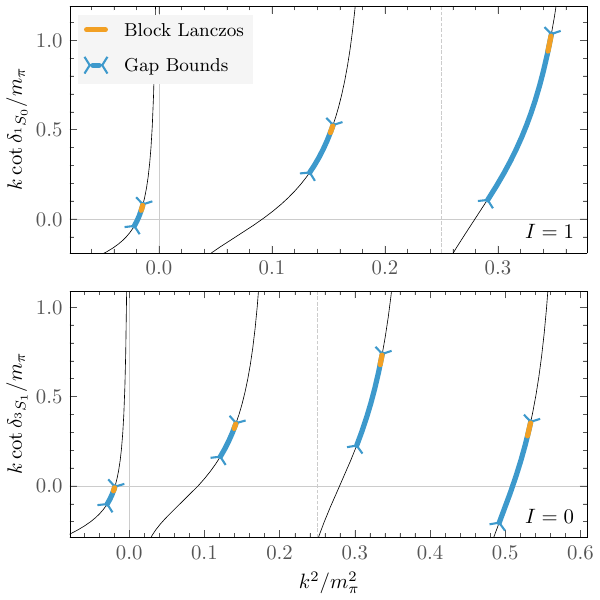}
    \caption{Scattering phase shift determinations from  the $S$-wave-truncated L\"uscher quantization condition~\cite{Luscher:1986pf,Luscher:1990ux} for  $I=1$ (top) and $I=0$ (bottom) channels. Orange bands denote statistical uncertainties. Chevron-bounded blue bands denote gap plus variational bounds at 68\% confidence.}
    \label{fig:phaseshifts}
\end{figure}

Given that $N$-state-saturation assumptions are known to fail for at least some operator sets in the $NN$ system, the use of gap bounds under less-constraining no-missing-states assumptions is well motivated. The gap parameter used for these bounds can be estimated from the non-interacting two-hadron spectrum. 
Using this prescription, consistent gap-bound constraints are obtained from all interpolator sets studied; sets with ``missing operators'' leading to significant discrepancies under $N$-state saturation assumptions simply lead to less precise gap bounds~\cite{long}.
The lowest energies obtained with one set of dibaryon operators are shown in \cref{tab:results} for both the scalar- and block-Lanczos results. With these assumptions, the two-sided bounds that are produced by combining gap plus variational bounds allow for the separation of different scattering states. Propagating these bounds through the L\"uscher quantization condition~\cite{Luscher:1986pf,Luscher:1990ux} leads to scattering phase-shift constraints shown in \cref{fig:phaseshifts}. It is apparent from this figure that gap bounds on energy eigenvalues at the level of statistics used in this work are sufficient for phenomenologically meaningful determinations of scattering phase shifts. 

The presence or absence of bound states in both two-nucleon channels at heavy quark masses has been debated in the literature~\cite{Fukugita:1994na,Fukugita:1994ve,Beane:2006mx,Ishii:2006ec,Beane:2009py,Yamazaki:2009ua,NPLQCD:2011naw,Murano:2011nz,Aoki:2011gt,NPLQCD:2012mex,Yamazaki:2012hi,Ishii:2012ssm,NPLQCD:2013bqy,Orginos:2015aya,Yamazaki:2015asa,Berkowitz:2015eaa,Wagman:2017tmp,Francis:2018qch,NPLQCD:2020ozd,NPLQCD:2020lxg,Horz:2020zvv,Amarasinghe:2021lqa,Detmold:2024iwz,BaSc:2025yhy}.
In principle, at the high level of statistics used here, variational or Lanczos bounds applied to correlator matrices could have validated the existence of a bound state at $\sim15$ MeV below threshold as has been suggested from asymmetric correlator 
analyses~\cite{NPLQCD:2013bqy,Wagman:2017tmp,Orginos:2015aya,NPLQCD:2020ozd,NPLQCD:2020lxg,Yamazaki:2012hi,Yamazaki:2015asa,Berkowitz:2015eaa}, but this was not the case. An oblique Lanczos analysis of the asymmetric correlators casts doubt on their physical interpretation and ability to provide positive evidence for a low-lying state; however, it is also important to emphasize that these results do not exclude such a state. The nature of the lowest energy state in these systems at these action parameters remains uncertain. 
Further discussion of this point and of adversarial models that explain the high-precision data of this work under both the bound- and scattering-state scenarios is presented in Ref.~\cite{long}.
One positive conclusion of this study at the level of the assumptions needed for gap bounds is that the $NN$ system at heavy quark masses is fine-tuned: the scattering lengths in both the $^1S_0$ and $^3S_1$ channels are unnaturally small with $|a^{-1}_{^{1}S_0/^{3}S_1}|\alt m_\pi/10$ at $68\%$ confidence, as seen in \cref{fig:phaseshifts}. 

Given the unphysically heavy quark masses used in this study, it is natural to ask whether these methods will be viable for physical values of the quark masses.
This question cannot yet be fully addressed, but Appendix~\ref{app:light}  presents
lower-statistics studies of the nucleon correlators for four close-to-physical quark-mass ensembles with varying lattice spacings, volumes, and masses corresponding to $m_{\pi} \in [135, 180]$ MeV.
They indicate that with present-day or near-future computing resources,  percent-level determinations of the nucleon mass  using either gap or residual bounds could be possible, which would bring spectroscopic uncertainties to a similar scale as discretization and finite volume effects.
More speculatively, it may be viable---albeit challenging---to apply these methods to physical-mass two-nucleon systems in future work.

{\it Summary:} In this work, the utility of two-sided bounds available in Lanczos approaches to correlator analysis for the study of one- and two-nucleon systems has been investigated through a high-statistics LQCD calculation at heavy quark masses. Spectral-assumption-free residual bounds provide sub-percent constraints on energies in one- and two-nucleon systems at heavy quark masses but do not currently provide sufficient constraints to enable extraction of scattering phase shifts. This latter failure provides motivation to develop improved methods, as new approaches to interpolating operator design may exist that dramatically reduce the sizes of the residual bounds in the Lanczos approach. At the same time, gap bounds based only on assumptions about the number of states in the spectrum within a specified window are found to be sufficiently precise to allow phenomenologically useful scattering phase shift constraints. They provide a more controlled approach to hadronic interactions than current methods that require $N$-state-saturation assumptions that are known to fail in some situations. The new tools presented here may also be useful in better quantifying uncertainties in correlator-based energy determinations in other fields.

\begin{acknowledgements}
We are grateful to R.~Brice{\~n}o, Z.~Davoudi, Y.~Fu, W.~Jay, C.~Morningstar, A.~Nicholson, A.~Parre{\~n}o, F.~Romero-L\'opez, M.~J.~Savage, and A.~Walker-Loud for helpful discussions.
The software packages chroma \cite{Edwards_2005}, qdp-jit \cite{Winter:2014dka}, quda \cite{Clark_2010}, QPhiX \cite{qphix}, and GLU \cite{glu} 
were used to generate gauge field configurations and compute quark propagators.
WD, PES and RJP are supported in part by the U.S. Department of Energy, Office of Science under grant Contract Number DE-SC0011090 and by the SciDAC5 award DE-SC0023116.
PES is additionally supported by the U.S. DOE Early Career Award DE-SC0021006 and by Simons Foundation grant 994314 (Simons Collaboration on Confinement and QCD Strings).
MI is partially supported by the Quantum Science Center (QSC), a National Quantum Information Science Research Center of the U.S. Department of Energy.
AVG is supported by the National Science Foundation under Grant No.~PHY-240227.
This document was prepared using the resources of the Fermi National Accelerator Laboratory (Fermilab), a U.S. Department of Energy, Office of Science, Office of High Energy Physics HEP User Facility. Fermilab is managed by Fermi Forward Discovery Group, LLC, acting under Contract No.~89243024CSC000002.

This research used resources of the Oak Ridge Leadership Computing Facility at the Oak Ridge National Laboratory, which is supported by the Office of Science of the U.S. Department of Energy under Contract number DE-AC05-00OR22725
and the resources of the National Energy Research Scientific Computing Center (NERSC), a Department of Energy Office of Science User Facility using NERSC award NP-ERCAPm747. 
We acknowledge EuroHPC JU for awarding the project ID EHPC-REG-2023R03-082 access to LUMI at CSC, Finland. 
The research reported in this work made use of computing facilities of the USQCD Collaboration, which are funded by the Office of Science of the U.S. Department of Energy.
\end{acknowledgements}

\bibliography{bib}

\crefalias{section}{appendix}

\clearpage

\appendix

\section{Nucleon spectroscopy at close-to-physical quark massses}\label{app:light}

To study how the precision of different estimators and bounds depends on quark masses, results for the nucleon are studied for a set of four close-to-physical quark mass ensembles summarized in \cref{tab:light_ensembles}; see Refs.~\cite{Yoon:2016jzj,Mondal:2021oot,Yoo:2026uul} for further details of ensemble generation.

\begin{table}[]
    \centering
    \begin{ruledtabular}
    \begin{tabular}{cccccccc}
      Ensemble & $L$ & $L_t$  & $\beta$ & $m_l$ & $m_s$ & $N_{\rm cfgs}$  \\\hline  
    A &  48 & 96
    & 6.3 & $-0.2416$ & $-0.2050$ & 5040 \\
    B &  64 & 128
    & 6.3 & $-0.2416$ & $-0.2050$ & 1422 \\
    C &  72 & 192 
    & 6.5 & $-0.2091$ & $-0.1778$ & 460 \\ 
     D & 96 & 192
    & 6.5 & $-0.2095$ & $-0.1793$ & 1352 \\
    \end{tabular}
    \end{ruledtabular}
    \caption{ Parameters of the close-to-physical quark-mass ensembles discussed in this section; see Refs.~\cite{Yoon:2016jzj,Mondal:2021oot,Yoo:2026uul}.
    }
    \label{tab:light_ensembles}
\end{table}

\begin{figure*}
    
    \includegraphics[width=0.48\textwidth]{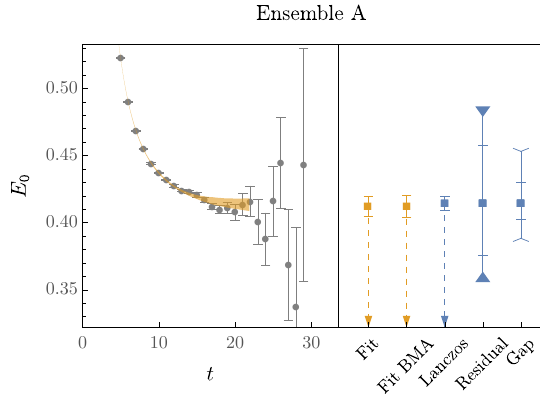} 
    \includegraphics[width=0.48\textwidth]{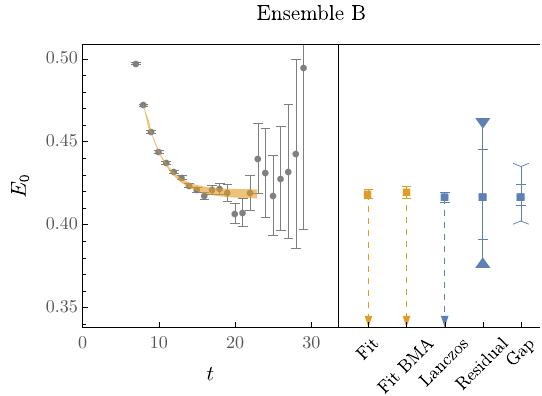} \\
    \includegraphics[width=0.48\textwidth]{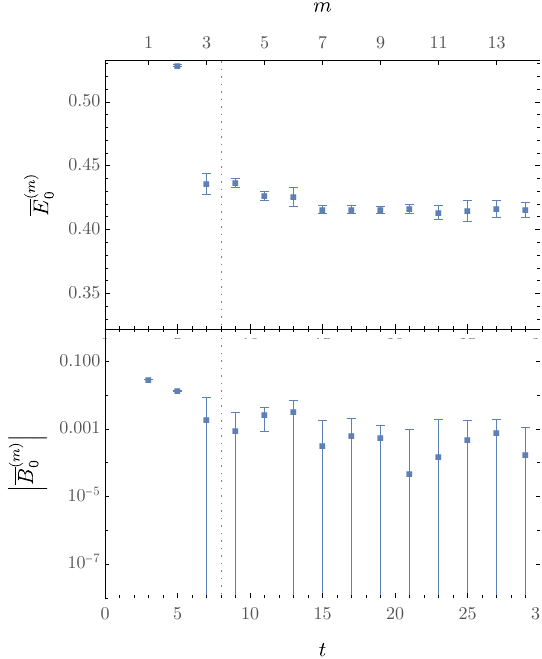}
    \includegraphics[width=0.48\textwidth]{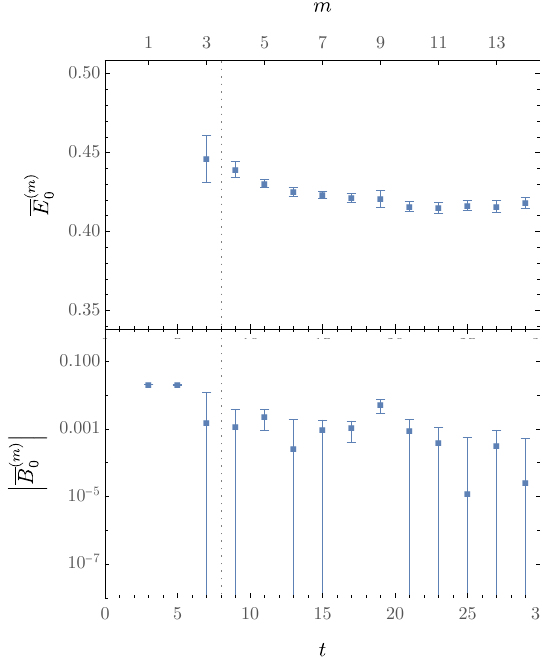}

    \caption{ Results for the $\beta=6.3$ ensemble with $m_{\pi} \approx 180$ MeV and $L=48$, left, and the same parameters with $L=64$, right.
    Upper: Nucleon effective masses and multi-state fit results (orange) along with Lanczos energy estimators and two-sided residual and gap bounds from a median of large-$m$ results (blue); details are as in \cref{fig:nn_summary}. Lower: Iteration dependence of Lanczos energy and residual-norm-square estimators. 
      A vertical dashed lines is placed after the median $m$ where spurious eigenvalues first appear; the previous iteration (at the inner bootstrap level) is used for SLRVL state identification; see Ref.~\cite{long}.
     Correlator results with $t \in \{0,1\}$ are omitted to avoid complications arising from contact terms. 
     \label{fig:nuc-EMP-64-48}
     }
\end{figure*}

\begin{figure*}
    \includegraphics[width=0.48\textwidth]{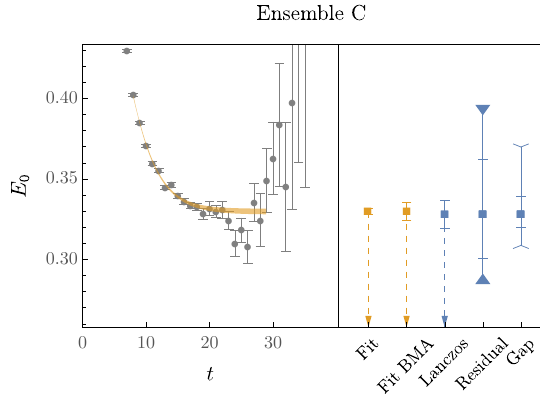} 
    \includegraphics[width=0.48\textwidth]{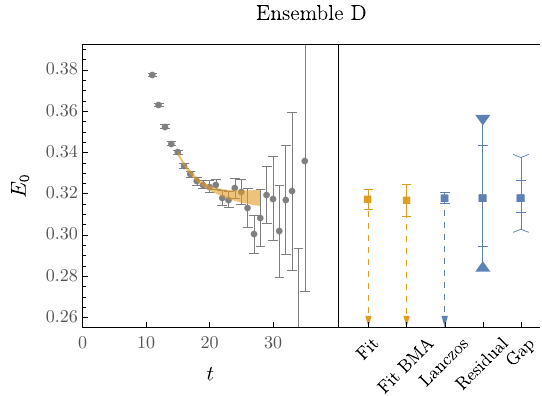} \\
    \includegraphics[width=0.48\textwidth]{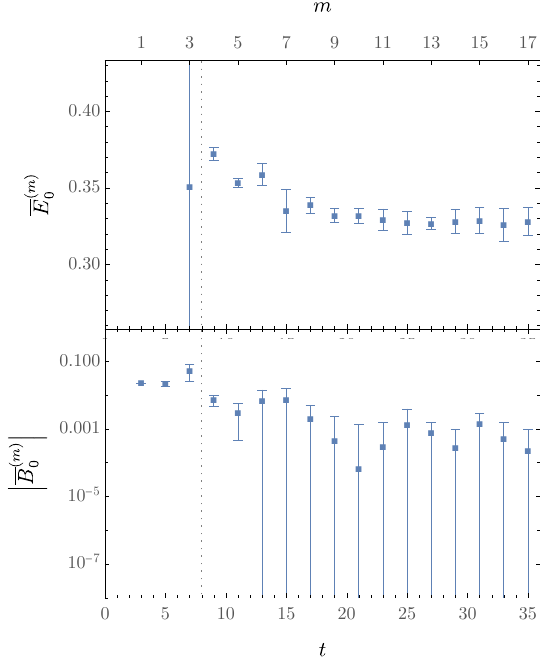}
    \includegraphics[width=0.48\textwidth]{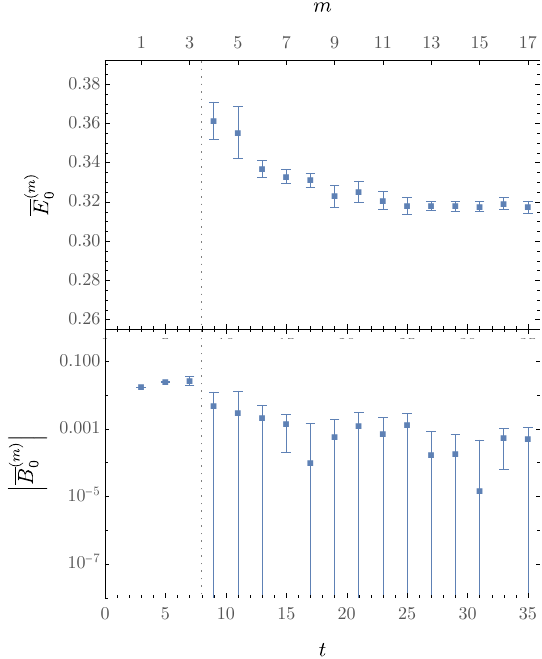}
    \caption{ Results for the $\beta=6.5$ ensemble with $m_{\pi} \approx 135$ MeV and $L=96$, left, and the $\beta=6.5$ ensemble with $m_{\pi} \approx 180$ MeV and $L=72$, right.
         Details are as in \cref{fig:nuc-EMP-64-48}.
     \label{fig:nuc-EMP-96-72}
     }
\end{figure*}

Nucleon correlator analysis can be performed just as in the main text and Ref.~\cite{long}.
In particular, all Lanczos results include Hermitian-subspace and ZCW filtering with $F_{ZCW}=10$ and SLRVL state identification~\cite{long}.
Results for the nucleon Lanczos ground-state energy estimators median-averaged over iterations $m\in \{9,\ldots,14\}$ for Ensembles A--B and $m \in \{11,\ldots,17\}$ for Ensembles C--D, denoted $\overline{E}_0^{(N)}$, are shown in \cref{fig:nuc-EMP-96-72,fig:nuc-EMP-64-48} and \cref{tab:light_results}.
Under $N$-state saturation assumptions, the results for all four ensembles are consistent at $1\sigma$ with corresponding energy estimates from Ref.~\cite{Yoo:2026uul}.

Two-sided bounds on the nucleon spectrum are also presented in \cref{tab:light_results}.
Gap bounds are computed with no-missing-states assumptions where
  \begin{align}
    \hat{E}_{\rm miss} &= \min \bigl\{ M_N + 2m_\pi,  \\\nonumber
    &\hspace{30pt} \sqrt{ M_N^2 + (2\pi/L)^2} + \sqrt{ m_\pi^2 + (2\pi/L)^2}  \bigr\},
  \end{align}
with $M_N$ estimated as $\overline{E}_0^{(N)}$ and $m_\pi$ estimated from the corresponding Lanczos energy estimator for the pion,\footnote{Energy estimators show 2--3$\sigma$ discrepancies with higher-statistics results from Ref.~\cite{Yoo:2026uul}, but gap bounds show $1\sigma$ consistency with those results under no-missing-states assumptions.} $\overline{E}_0^{(\pi)}$, with results shown in \cref{tab:light_results_pi}. Analogous results for the ratio $M_N/m_\pi$, for which two-sided bounds involve appropriate ratios of upper (lower) bounds on $E_0^{(N)}$ and lower (upper) bounds on $E_0^{(\pi)}$, are shown in \cref{tab:light_results_ratio}.

Results computed using 500--5000 gauge-field configurations, corresponding to $10^5$--$10^6$ quark propagator sources, achieve percent-level statistical uncertainties on $\overline{E}_0^{(N)}$: precision across the four ensembles ranges from 0.8\%--2.6\%.
Gap bounds achieve 3\%--6\% precision.
Residual bounds achieve 9\%--14\% precision.
For all four ensembles, $\overline{B}_0^{(N)}$ is consistent with zero within $1\sigma$ with magnitude ranging from 2--5 times $10^{-4}$.
Increasing statistics can be expected to lead to approximately $1/\sqrt{N_{\rm cfgs}}$ reduction in the magnitude and uncertainties of $\overline{B}_0^{(N)}$ as discussed in  Ref.~\cite{long}.

It is noteworthy that the ratio of 68\% confidence residual-bound width to statistical uncertainty, $\approx 9$--13, is much smaller for these close-to-physical quark-mass results than the corresponding ratio, $\approx 43$, for the $m_\pi \sim 800$ MeV results discussed in the main text.\footnote{Note that residual-bound width estimators are highly non-Gaussian. The ratio of 95\% confidence residual-bound width to $2\sigma$ statistical uncertainty is  $\approx 6$--9 for close-to-physical quark-mass results.}
A similar feature is seen for $\overline{E}_0^{(\pi)}$, where $68\%$ residual-bound widths are $\approx 28$--37 times larger than statistical uncertainties for close-to-physical quark-mass results while $\approx 90$ times larger for $m_\pi \approx 800$ MeV.

On the other hand, gap bounds are relatively less constraining for lighter quark masses.
The ratio of 68\% confidence gap-bound width to statistical uncertainty, $\approx 4$--6, is much larger for these close-to-physical quark-mass results than the corresponding ratio, $\approx 1.5$, for the $m_\pi \sim 800$ MeV results.
This could be explained in part by the fact that the relative size of the gap parameter $(\lambda_0^{(N)} - \lambda_1^{(N)})/\lambda_0^{(N)}$ is $\approx 3$--5 times smaller for these $m_\pi \lesssim 180$ MeV results than for $m_\pi \sim 800$ MeV results. 
More sophisticated calculations including explicit $N\pi$ and $N\pi\pi$ operators might therefore improve this situation by enabling increased values of $\hat{E}_{\rm miss}$ to be used in gap bounds, although the empirical growth of residual-norm-squares with correlator matrix rank observed in Refs.~\cite{Hackett:2024nbe,long} could wash out these gains.

The results of these close-to-physical quark-mass studies indicate that constraints on hadron masses with few-percent precision can be obtained from both gap and residual bounds using statistical ensembles that are larger than the ones considered here but plausibly within the reach of dedicated campaigns using present-day or near-future computing resources.
The level of precision required of spectroscopy results of course depends on the application---lattice-QCD results will not compete with experimental precision on the nucleon mass in the foreseeable future under even the most constraining assumptions---but this demonstrates a path for spectroscopy results that do not assume excited-state effects are small to achieve a level of precision where other systematic uncertainties related to discretization, finite volume, quark-mass mistuning, isospin breaking, and QED become dominant.

\begin{table*}[h!]
    \centering
    \begin{ruledtabular}
    \begin{tabular}{cccccc}
      Ensemble & $\overline{E}_0^{(N)}$  & Residual (68\%)&  Residual (95\%)&  Gap + variational (68\%) &Gap + variational (95\%) \\\hline  
          A 
    & $0.415(5)_{(-26)}^{(+0)}$ 
      & $[0.36, 0.48]$ & $[0.33, 0.52]$ 
       & $[0.388, 0.420]$ & $[0.356, 0.425]$ \\
        B 
    & $0.416(3)_{(-14)}^{(+0)}$  
         & $[0.38, 0.46]$ & $[0.36, 0.48]$ 
       & $[0.402, 0.419]$ & $[0.388, 0.423]$ \\
    C 
    & $0.328(9)_{(-19)}^{(+0)}$  
     & $[0.29, 0.39]$ & $[0.27, 0.49]$ 
      & $[0.309, 0.337]$ & $[0.232, 0.345]$ \\ 
      D 
    & $0.318(3)_{(-15)}^{(+0)}$ 
      & $[0.28, 0.36]$ &  $[0.27, 0.38]$ 
       & $[0.303,0.321]$ & $[0.286,0.323]$ 
    \end{tabular}
    \end{ruledtabular}
    \caption{ Constraints on the nucleon mass from various close-to-physical quark-mass ensembles. The second column shows ground-state energies with 68\% statistical uncertainties followed by a ``systematic uncertainty'' that when added to the statistical uncertainty in quadrature produces the widths of combined variational and Haas-Nakatsukasa gap bounds at 68\% confidence.  The third (fifth) columns shows residual bounds at 68\% (95\%) confidence computed using the $k D$ interpolator set, which in all cases are tighter than bounds obtained from larger interpolator sets. The fourth (sixth) column shows gap bounds at 68\% (95\%) confidence. All gap bounds results are computed with the minimum of the non-interacting $N\pi$ $P$-wave and $N\pi\pi$ $S$-wave energies used to estimate the gap parameter.
    }
    \label{tab:light_results}
\end{table*}

\begin{table*}[h]
    \centering
    \begin{ruledtabular}
    \begin{tabular}{cccccc}
      Ensemble & $\overline{E}_0^{(\pi)}$  & Residual (68\%)&  Residual (95\%)&  Gap (68\%) &Gap (95\%) \\\hline  
    A 
    & $0.0760(6)_{(-50)}^{(+54)}$ 
      & $[0.057, 0.095]$ & $[0.047, 0.100]$ 
       & $[0.0710, 0.0814]$ & $[0.0644, 0.0964]$ \\
       B 
    & $0.0777(2)_{(-9)}^{(+9)}$  
    & $[0.070, 0.085]$ & $[0.064, 0.091]$ 
       & $[0.0768, 0.0786]$ & $[0.0752, 0.0831]$ \\
    C 
    & $0.0587(5)_{(-67)}^{(+68)}$  
    & $[0.040, 0.078]$ & $[0.032, 0.089]$ 
      & $[0.0523, 0.0658]$ & $[0.0453, 0.0901]$ \\
       D 
    & $0.0457(2)_{(-8)}^{(+9)}$ 
    & $[0.040, 0.052]$ &  $[0.036, 0.055]$ 
       & $[0.0448,0.0466]$ & $[0.0432,0.0498]$ 
    \end{tabular}
    \end{ruledtabular}
    \caption{ Constraints on the pion mass from various close-to-physical quark-mass ensembles. Note that thermal effects are significant for the pion and therefore energy estimators should not be interpreted as variational bounds. Other details are as in \cref{tab:light_results}.
    }
    \label{tab:light_results_pi}
\end{table*}

\begin{table*}[h]
    \centering
    \begin{ruledtabular}
    \begin{tabular}{cccccc}
      Ensemble & $\overline{E}_0^{(N)} / \overline{E}_0^{(\pi)}$  & Residual (68\%)&  Residual (95\%)&  Gap + variational (68\%) &Gap + variational (95\%) \\\hline  
    A 
    & $5.45(9)_{(-18)}^{(+18)}$ 
      & $[3.8, 8.3]$ & $[3.5, 10]$ 
       & $[4.84, 5.90]$ & $[4.39, 6.59]$ \\ 
        B 
    & $5.36(5)_{(-10)}^{(+6)}$  
      & $[4.5, 6.5]$ & $[4.1, 7.2]$ 
       & $[5.14, 5.47]$ & $[4.97, 5.57]$ \\
         C 
    & $5.59(17)_{(-32)}^{(+42)}$  
      & $[3.8, 9.5]$ & $[3.4, 12]$ 
      & $[4.75, 6.54]$ & $[3.83, 7.75]$ \\
       D 
    & $6.96(6)_{(-18)}^{(+9)}$ 
      & $[5.6, 8.8]$ &  $[5.1, 9.8]$ 
       & $[6.50, 7.17]$ & $[6.17, 7.44]$ \\
    \end{tabular}
    \end{ruledtabular}
    \caption{ Constraints on the ratio $M_N / m_\pi$ from various close-to-physical quark-mass ensembles. Details are as in \cref{tab:light_results}.
    }
    \label{tab:light_results_ratio}
\end{table*}

\end{document}